\documentclass[aps,twocolumn,pra,showpacs]{revtex4}
\usepackage{amssymb}
\usepackage{amsmath}
\usepackage{graphicx}
\usepackage{epsfig}
\usepackage{subfigure}
\usepackage{amsfonts}
\usepackage{CJK}
\usepackage{color}

\begin{document}

\title{Creation, manipulation and detection of Majorana fermions with cold
atoms in optical lattice}
\author{Feng Mei$^{1,3}$}
\author{Chuan-Jia Shan$^{2}$}
\author{Xun-Li Feng$^{3}$}
\author{ Shi-Liang Zhu$^{2}$}
\email{shilzhu@yahoo.com.cn}
\author{Zhi-Ming Zhang$^{1}$}
\email{zmzhang@scnu.edu.cn}
\author{L. C. Kwek$^{3,4}$}
\email{kwekleongchuan@nus.edu.sg}
\author{D. Wilkowski$^{3,5,6}$}
\email{david.wilkowski@ntu.edu.sg}
\author{C. H. Oh$^{3}$}
\email{phyohch@nus.edu.sg}
\affiliation{$^{1}$Laboratory of Nanophotonic Functional Materials and Devices, LQIT $\&$
SIPSE, South China Normal University, Guangzhou 510006, China\\
$^{2}$Laboratory of Quantum Information Technology and SPTE, South China
Normal University, Guangzhou, China\\
$^{3}$Centre for Quantum Technologies and Department of Physics, National
University of Singapore, 3 Science Drive 2, Singapore 117543, Singapore\\
$^{4}$National Institute of Education and Institute of Advanced Studies,
Nanyang Technological University, 1 Nanyang Walk, Singapore \\
$^{5}$Institut Non Lin\'eaire de Nice, Universit\'e de Nice
Sophia-Antipolis, CNRS, 06560 Valbonne, France\\
$^{6}$School of Physical and Mathematical Sciences, Nanyang Technological
University, Singapore 637371, Singapore}

\begin{abstract}
We propose an experimental scheme to simulate the transverse field Ising
model with cold atoms trapped in one-dimensional optical lattice. Majorana
fermions are created at the ends of the optical lattice segment in
topological phase. By controlling the addressing lasers, one can move, fuse
and braid them. We also show that the non-Abelian braiding statistics of
Majorana fermions can be demonstrated unambiguously through the construction
of two braiding operations and distinguishing the resulting two output
orthogonal collective spin states. A nice feature of the scheme is that the
strong fluorescence provided by the collective spin state can be readily
detected in experiment.
\end{abstract}

\maketitle


Majorana fermions are particles that are their own antiparticles unlike
Dirac fermions where electrons and positrons are distinct {\color{blue}\cite%
{Wilczek}}. When exchanged among themselves, Majorana fermions obey
non-Abelian statistics. Such fundamentally interesting particles have
recently attracted much attention due to their potential application in
topological quantum computation {\color{blue}\cite{TQC}}. Majorana modes are
originally perceived as zero-energy states bound to the vortices in
two-dimensional (2D) spinless $p_{x}\pm ip_{y}$-wave superconductor (SPSC) {%
\color{blue}\cite{Read}} or the two ends in one-dimensional (1D) SPSC chain {%
\color{blue}\cite{Kitaev01}}. It has been recently proposed that a
semiconductor thin film with Rashba spin-orbit coupling, together with
proximity-induced superconductivity and Zeeman splitting, resembling the
SPSC model, can be used to create Majorana fermions {\color{blue}\cite{2DMF}}%
. One dimensional version of this system has been shown to host Majorana
fermions at the two ends of a semiconducting wire {\color{blue}\cite{1DMF}}.
By using a T-junction wire network, the Majorana fermions in the 1D
semiconducting wires can be braided by tuning the local gates {\color{blue}%
\cite{Alicea}}. Despite the relative ease with which Majorana fermions can
be stabilized in 1D wires making this 1D set-up more promising as a
topological quantum information processing platform, experimental challenges
need to be overcome: strong Zeeman fields could destroy the superconductor
and the local control of electron density by the gates is non-trival due to
strong screening by the superconductor.

Cold atoms nowadays are widely recognized as powerful experimental tools for
mimicking a wide range of systems originally stemming from condensed-matter
physics. This system can provide a highly controllable and tunable
environment. Many significant experimental advances have been made in this
forefront field. In particular, recent experiments have realized synthetic
magnetic fields{\ }and spin-orbit coupling for ultracold atoms {\color{blue}%
\cite{Spielman1,Spielman2}}. Several protocols have been put forward to use
these technologies for generating and probing Majorana fermions in 2D and 1D
fermionic gases {\color{blue}\cite{Zhu,Tewari,Santo,Liu,Jiang,Zoller}}.
However, the braiding Majorana fermions as well as the detection of their
non-Abelian statistics remains an outstanding problem.

In this work, we propose an experimental scheme to create and braid
Majorana fermions as well as to detect their non-Abelian statistics with
cold atoms in optical lattice. Note that the transverse field Ising model
(TFIM) after Jordan-Winger (JW) transformation is equivalent to the 1D SPSC
model {\color{blue}\cite{Kitaev09,You-Solano}}. We show that various TFIMs
can be simulated with two-component ultracold bosonic atoms trapped in a 1D
optical lattice. This atom-lattice system is highly tunable and very robust.
The spin exchange coupling and the transverse field can be tuned by changing
the laser intensity. Under this tuning, the optical lattice can be driven
into topological phase or non-topological phase. Majorana fermions are bound
to the ends of the optical lattice segment in topological phase. Such tuning
can also allow Majorana fermions to be moved, recreated and fused. By
further employing a cross lattice, Majorana fermions can be braided in the
1D optical lattice. In addition, the orientation of our simulated Ising
coupling can be conveniently rotated by varying the phase of the laser.
Finally, we show that such rotation can also mimic the braiding of Majorana
fermions.

Unambiguous detection of non-Abelian statistics of Majorana fermions through
easily measurable collective spin states is another distinct advantage of
our proposal. We use the braid group elements of Majorana fermions to
construct two different orders of braiding operations. When applied to the
same initial state, the corresponding output states are orthogonal. For 1D
superconductor model, the state of the non-local Majorana fermions has
neutral charge and it is hard to be detected. However, the Majorana fermion
states in our spin chain model can be mapped back to spin basis so that the
states of two output Majorana fermions are two orthogonal collective spin
states. Compared with the detection of the fluorescence from single atom {%
\color{blue}\cite{Zhu,Tewari}}, which would be hard, these collective states
consist of hundreds of atoms and they can provide strong fluorescence, which
allows for experimental detection with high efficiency. In this way, our
proposed scheme provides an easy way to detect the fundamental non-Abelian
statistics of Majorana fermions.

Let us consider an ensemble of ultracold $^{87}$Rb atoms trapped in a 1D
optical lattice at $1064$~nm. With standard optical pumping methods, we
suppose that the atoms can only be in the states $|5S_{1/2},F=2,m_{F}=1%
\rangle $ or $|5S_{1/2},F=1,m_{F}=0\rangle $ denoted in the following text
with an effective spin index $\sigma =\downarrow $, $\uparrow $. Taking into
account the ground state hyperfine splitting $\Delta \nu _{hf}=6.8$~GHz, the
AC polarizabilities for the spin up and spin down have a small difference of
about $3\%$. As shown in Fig. 1, it can be used to generate a
state-independent optical lattice by applying a standing wave laser beam $%
L_{1}$. By considering a trapping depth of $20E_{r}$, where $E_{r}$ is the
lattice recoil energy, the 1D potential is almost state-independent within a
mismatch on the hopping rate lower than $15\%$. In order to simulate the
Ising model in this 1D optical lattice, we need to apply a second
state-dependent optical lattice with a potential $V_{\sigma }$. As we will
see, this state-dependent potential is very important for realizing the
Ising model.

At sufficiently low temperatures and strong lattice potential, the atoms are
confined to the ground state of the optical lattice, and the system can be
described by a Bose-Hubbard Hamiltonian
\begin{equation}
H=-\sum\limits_{<i,j>\sigma }\left( t_{\sigma }a_{i\sigma }^{+}a_{j\sigma
}+H.c\right) +\frac{1}{2}\sum\limits_{i,\sigma \sigma ^{\prime }}U_{\sigma
\sigma ^{\prime }}\colon n_{i\sigma }n_{i\sigma ^{\prime }}\colon \text{,}
\end{equation}%
where $a_{i\sigma }$ is the bosonic annihilation operator for atom states of
spin $\sigma $ at the lattice site $i$, $n_{i\sigma }=a_{i\sigma
}^{+}a_{i\sigma }$, $\colon (...)\colon $denotes normal order of the product
of creation-annihilation operators, $t_{\sigma }$ is the spin-dependent
hopping rate and $U_{\sigma \sigma ^{\prime }}$ is the on-site energy
between species $\sigma $ and $\sigma ^{\prime }$. It is known that $%
t_{\sigma }$ depends sensitively upon the lattice potential while $U_{\sigma
\sigma ^{\prime }}$ exhibits weak dependence {\color{blue}\cite{Duan}}.
Moreover, for the state under consideration the scattering lengths are $%
a_{\uparrow \uparrow }\simeq a_{\downarrow \downarrow }\simeq a_{\downarrow
\uparrow }$ . Thus, in what follows, $U_{\sigma \sigma ^{\prime }}=U$ is
assumed. By varying the intensity of trapping laser to change the lattice
potential, one can control the ratio $U/t_{\sigma }$ to realize the Mott
insulator with unitary filling {\color{blue}\cite{Mott02}}, we will now
consider this case in our model.

\begin{figure}[tbp]
\includegraphics[scale=0.5]{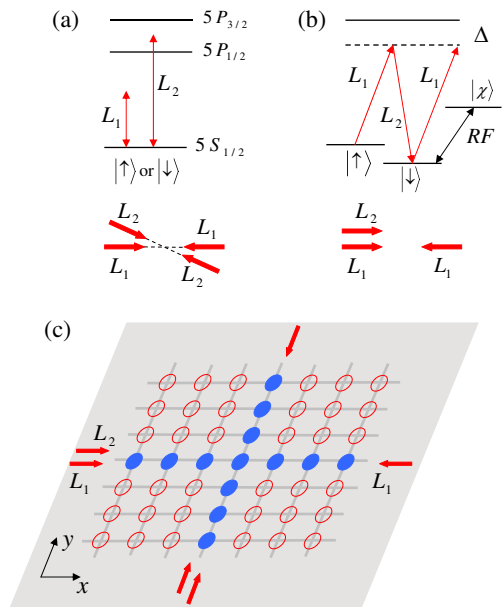}
\caption{(Color online) Energy diagram and laser beam configurations for the
$\protect\sigma ^{z}$ (a) and $\protect\sigma ^{\protect\varphi }$ (b) Ising
models. $L_{1}$ and $L_{2}$ are the lasers leading to the 1D
state-independent optical lattice and spin-dependent hopping rate. (c) 2D
lattice for the braiding of Majorana fermion. The filled blue (open red)
circles symbolized the atoms in the $|\protect\sigma \rangle $ ($|\protect%
\chi \rangle $) state. An $RF$-field is supposed to be used to flip the
state in $|\protect\chi \rangle $ to the spin state $|\protect\sigma \rangle
$.}
\end{figure}

By utilizing an additional state-dependent potential, the above Hubbard
Hamiltonian is reduced to the $\sigma ^{z}$- and $\sigma ^{\varphi }$-Ising
models. For $\sigma ^{z}$-Ising model, as sketched in Fig. 1(a), we apply a
weak standing-wave laser beam $L_{2}$, detuned between the D2 ($780$~nm) and
D1 ($794$~nm) lines. In order to make complete overlap with the first
state-independent optical lattice, the standing-wave laser beam should have
an proper angle of incidence to the first one. If one tunes the intensity of
the laser beam so that $V_{\uparrow }\gg V_{\downarrow }$ (or vice-versa),
the hopping rate $t_{\uparrow }$ becomes negligible while $t_{\downarrow }$
remains finite, after compensating effective B-fields, the effective
Hamiltonian of Eq. (1) could be reduced to the $\sigma ^{z}$-Ising model $%
J_{1}\sum_{j=1}^{N-1}\sigma _{j}^{z}\sigma _{j+1}^{z}$ with $%
J_{1}=-t_{\downarrow }^{2}/2U$ {\color{blue}\cite{Duan}}. This method
requires an interferometric control of the spatial overlapping of the two
standing waves at the atomic cloud position, which may be experimentally
challenging.

An alternative and simpler method, for $\sigma ^{\varphi }$-Ising model,
consists of coupling the effective spin states via a two-photon Raman
process. One photon is issued from the 1D lattice beams with a Rabi
frequency $\Omega _{1}\cos ^{2}{kx}$, where $k$ is the wave number of the
laser. The second photon comes from a running wave, $L_{2}$, propagating
along $x>0$ with a complex Rabi frequency $\Omega _{2}e^{i\varphi }$ (see
Fig. 1(b)). At the equilibrium position of the atoms, the complex Rabi
frequency of the Raman coupling is $\Omega _{R}=\Omega _{1}\Omega
_{2}e^{i\varphi }/\Delta $, where $\Delta $ is the one photon detuning. One
notes that $\varphi $ can be controlled by standard phase lock techniques
and adjusted over a temporal delay of $(4\pi \Delta \nu _{hf})^{-1}\simeq 12$%
~ps. At resonance, the Raman coupling generates the state-dependent dressed
potential $V_{2}=V_{+}\left\vert +\right\rangle \left\langle +\right\vert
+V_{-}\left\vert -\right\rangle \left\langle -\right\vert $, where $%
\left\vert \pm \right\rangle =(\left\vert \uparrow \right\rangle \pm
e^{i\varphi }\left\vert \downarrow \right\rangle )/\sqrt{2}$, $%
V_{+}=-V_{-}=\hbar \left\vert \Omega _{R}\right\vert /2$. The total
state-dependent potential barriers now become $\tilde{V}_{s}=$ $V_{0}+V_{s}$
($s=\pm $). By improving the intensity of laser beam $L_{2}$ to make $\tilde{%
V}_{+}\gg \tilde{V}_{-}$, the hopping rate $t_{+\text{ }}$for the atom in
the dressed state $\left\vert +\right\rangle $ becomes negligible. This can
be achieved with moderate laser power. In this case, the effective
Hamiltonian of Eq. (1) can be rewritten as the $\sigma ^{\varphi }$-Ising
model $J_{2}\sum_{j=1}^{N-1}\sigma _{j}^{\varphi }\sigma _{j+1}^{\varphi },$
where $J_{2}=-t_{-}^{2}/2U$ and $\sigma ^{\varphi }=\cos \varphi \sigma
^{x}+\sin \varphi \sigma ^{y}$.

To realize the TFIM, we still need to generate the transverse field. For
transverse field $\sigma ^{z}$-Ising model, an addressing laser beam with
Rabi frequency $\Omega _{x}$ is applied to the lattice sites, which can
generate the transverse field term $h_{x}\sigma _{j}^{x}$. Combined with the
Ising Hamiltonian, we realize the transverse field $\sigma ^{z}$-Ising model

\begin{equation}
H_{1}=J_{1}\sum_{j=1}^{N-1}\sigma _{j}^{z}\sigma
_{j+1}^{z}+h_{x}\sum_{j=1}^{N}\sigma _{j}^{x},
\end{equation}%
where the transverse field $h_{x}=\hbar \Omega _{x}/2$. For transverse field
$\sigma ^{\varphi }$-Ising model, the transverse magnetic field $h_{z}\sigma
_{j}^{z}$ can be generated by applying an external magnetic field or by
detuning the 2 photons Raman. We get a state-dependent AC shift $V_{z}\sigma
_{j}^{z}$, which is equivalent to the effective transverse magnetic field.
Thus we arrive at the transverse field $\sigma ^{\varphi }$-Ising model

\begin{equation}
H_{2}=J_{2}\sum_{j=1}^{N-1}\sigma _{j}^{\varphi }\sigma _{j+1}^{\varphi
}+h_{z}\sum_{j=1}^{N}\sigma _{j}^{z},
\end{equation}%
where the transverse magnetic field $h_{z}=V_{z}$. In particular, when the
laser phase $\varphi $ is tuned to $0$ or $\pi /2$, the above model is with
respect to the transverse field $\sigma ^{x}$- or $\sigma ^{y}$-Ising model.
As one can see, our simulated TFIM is highly tunable based on controlling
the lasers, including tuning the orientations of Ising interaction, the spin
exchange coupling and the transverse magnetic field.

Kitaev has shown that the transverse field $\sigma ^{x}$-Ising model after a
JW transformation is equivalent to the 1D SPSC model with superconducting
phase $\theta =0$ {\color{blue}\cite{Kitaev09}}. This result also holds true
for transverse field $\sigma ^{y}$- and $\sigma ^{z}$-Ising models. Here we
further show that our simulated transverse field $\sigma ^{\varphi }$-Ising
model can be reduced to 1D SPSC model with tunable superconducting phase.
For our purpose, through employing the string-like annihilation and creation
operator $a_{j}=\sigma _{j}^{-}\prod_{i=1}^{j-1}\sigma _{i}^{z}$ and $%
a_{j}^{+}=\sigma _{j}^{+}\prod_{i=1}^{j-1}\sigma _{i}^{z}$, the transverse
field $\sigma ^{\varphi }$-Ising model can be rewritten as

\begin{eqnarray}
H_{2} &=&J_{2}\sum_{j=1}^{N-1}(e^{i\varphi }a_{j}-e^{-i\varphi
}a_{j}^{+})(e^{i\varphi }a_{j+1}+e^{-i\varphi }a_{j+1}^{+})  \notag \\
&&+h_{z}\sum_{j=1}^{N}(2a_{j}^{+}a_{j}-1),
\end{eqnarray}%
where the Dirac fermions satisfy the anticommutation relationship $%
\{a_{i},a_{j}^{+}\}=\delta _{ij}$. By associating the spin exchange coupling
$J_{2}$ to the hopping amplitude and the superconducting gap, the laser
phase $\varphi $ to the superconducting phase and the magnetic field $h_{z}$
to the chemical potential, the transverse field $\sigma ^{\varphi }$-Ising
model is mapped into the 1D SPSC model with tunable superconducting phase.
Based on the $N$ Dirac fermions annihilation and creation operators, $2N$
Majorana fermions operators are defined as
\begin{equation}
\gamma _{A,j}=e^{i\varphi }a_{j}+e^{-i\varphi }a_{j}^{+}\text{, \ }\gamma
_{B,j}=i(e^{-i\varphi }a_{j}^{+}-e^{i\varphi }a_{j}),
\end{equation}%
where $\gamma _{\alpha ,j}^{2}=1$ and $\{\gamma _{\alpha ,j},\gamma _{\beta
,k}^{+}\}=2\delta _{\alpha \beta }\delta _{jk}$, $\alpha $, $\beta =A$, $B$.
In terms of these operators, the Hamiltonian in Eq. (4) becomes

\begin{equation}
H_{2}=iJ_{2}\sum_{j=1}^{N-1}\gamma _{B,j}\gamma
_{A,j+1}+ih_{z}\sum_{j=1}^{N}\gamma _{A,j}\gamma _{B,j}.
\end{equation}%
When $J_{2}\neq 0$ and $h_{z}=0$, the system is in the topological phase
with two unpaired Majorana end modes $\gamma _{A,1}$ and $\gamma _{B,N}$,
while for $J_{2}=0$ and $h_{z}\neq 0$, there is no Majorana end modes and
the lattice is in the non-topological phase. In the previous case, the two
Majorana fermions can be combined into an ordinary non-local Dirac fermion $%
c=(\gamma _{A,1}+i\gamma _{B,N})/2$, which yields two degenerate ground
states $\left\vert 0\right\rangle $ and $\left\vert 1\right\rangle
=c^{+}\left\vert 0\right\rangle $ {\color{blue}\cite{Kitaev01}}. Because the
fermion parity $P=-i\prod_{j=1}^{N}\gamma _{A,j}\gamma _{B,j}$ has a Z$_{2}$
symmetry, the ground states $\left\vert 0\right\rangle $ and $\left\vert
1\right\rangle $ have even and odd parity, i.e. $P\left\vert 0\right\rangle
(\left\vert 1\right\rangle )=\left\vert 0\right\rangle (-\left\vert
1\right\rangle )$. Due to the nonlocality of Majorana fermions that protects
the ground states from decoherence, one can use them to encode a topological
qubit for topological quantum memory.
\begin{figure}[tbp]
\includegraphics[width=8cm,height=5cm]{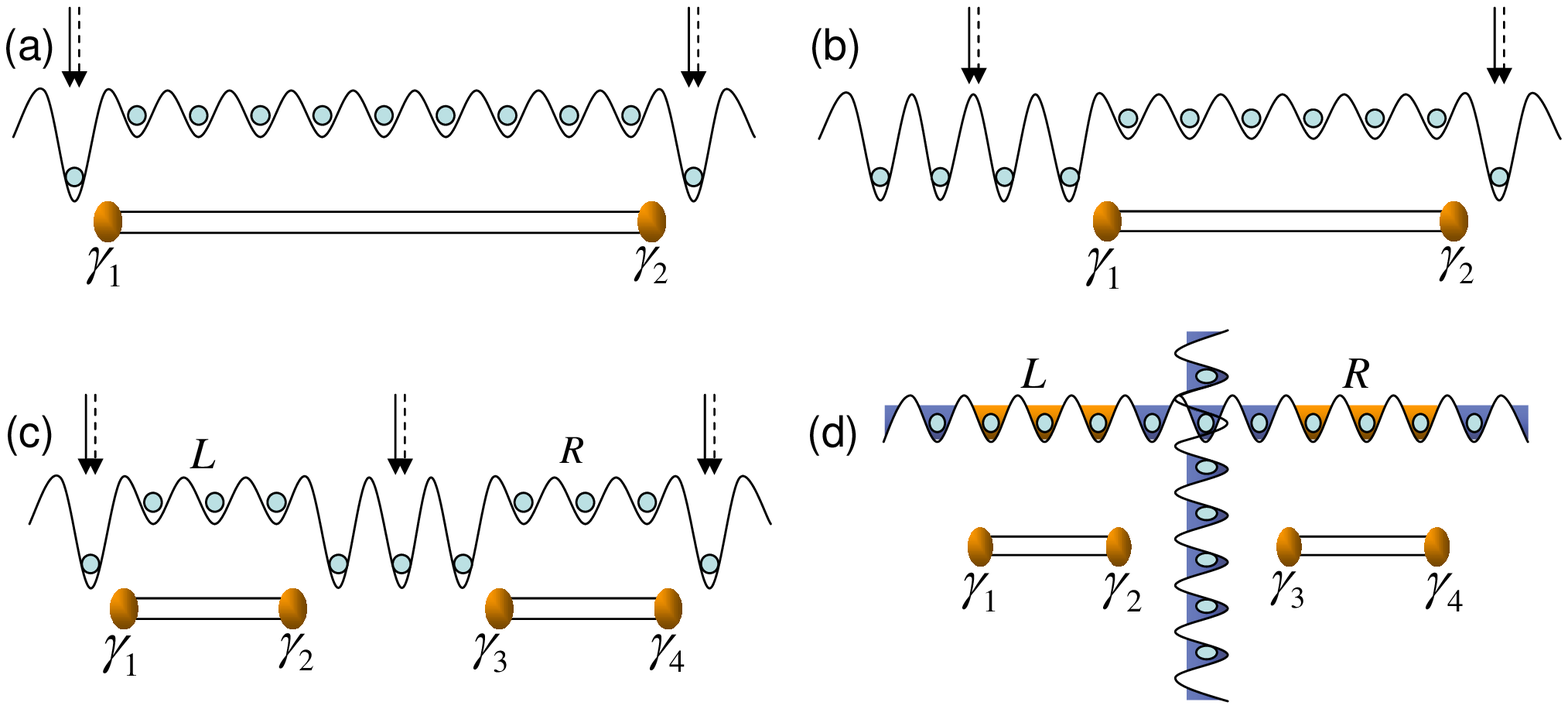}
\caption{(Color online) Schematic setup for manipulating and braiding
Majorana fermions. By controlling the addressing lasers, one pair of
Majorana fermions $\left( \protect\gamma _{1}\text{, }\protect\gamma %
_{2}\right) $ is created and moved in (a) and (b). With the same control,
two optical lattice segments $L$ and $R$ are driven into topological phase
and two pairs of Majorana fermions $\left( \protect\gamma _{1}\text{, }%
\protect\gamma _{2}\right) $ and $\left( \protect\gamma _{3}\text{, }\protect%
\gamma _{4}\right) $ are generated in (c). To braid these Majorana fermions,
a cross lattice is introduced in (d), where the orange segment denotes the
topological phase while the blue one denotes non-topological. }
\end{figure}

The spin exchange coupling $J_{2}$ and the transverse field $h_{z}$ can be
tuned by controlling the addressing laser beams. As displayed in Fig.
2(a-b), such tuning allows Majorana fermions to be created, moved and fused
in the 1D optical lattice. In Fig. 2(a), we assume that the lattice is a
simulated transverse field $\sigma ^{\varphi }$-Ising model and we can
address its two edge sites with two laser beams, one for improving the local
state-independent potential to make $J_{2}=0$ and one for generating the
transverse field $h_{z}$. The two edge lattice segments are driven into
non-topological phase but the center segment remains in the topological
phase, which leads to one pair of Majorana fermions $\gamma _{1}$ and $%
\gamma _{2}$ created at the ends of the topological segments. In Fig. 2(b),
by sequentially driving the leftmost lattice segment non-topological, the
Majorana fermion $\gamma _{1}$ is moved rightward and finally fused when it
meets the Majorana fermion $\gamma _{2}$. The experimental requirement of
single site addressing is now possible with high resolution microscope {%
\color{blue}\cite{Bakr,Sherson}}. As we will see later, this manipulation
constitutes basic operations needed to exchange Majorana fermions and
realize their non-Abelian braiding statistics.

Another feature of our model is that the superconducting phase for JW
fermions is tunable. Interestingly, we find that this tuning can be used to
simulate the braiding of the Majorana fermions at the ends of the same
topological segment. As described in Fig. 2(a), two Majorana fermions $%
\gamma _{A,1}$ and $\gamma _{B,N}$ are located at the ends of the center
topological lattice segment. According to Eq. (5), they can be written as $%
\gamma _{A,1}=e^{i\varphi }c+e^{-i\varphi }c^{+}$, $\gamma
_{B,N}=i(e^{-i\varphi }c^{+}-e^{i\varphi }c)$. By adiabatically tuning the
laser phase from $\varphi $ to $\varphi \pm \pi /2$, we get the following
transformation $\gamma _{A,1}\rightarrow \mp \gamma _{B,N}$, $\gamma
_{B,N}\rightarrow \pm \gamma _{A,1}$, which just implements the braiding
operation $\tau =\exp (\mp \frac{\pi }{4}\gamma _{B,N}\gamma _{A,1})=\exp
(\mp \frac{\pi }{4}\tilde{\sigma}_{z})$ {\color{blue}\cite{Alicea}}, where $%
\tilde{\sigma}_{z}$ is a Pauli matrix in the fermionic occupation basis $%
\left\{ \left\vert 0\right\rangle \text{, }c^{+}\left\vert 0\right\rangle
\right\} $. Moreover, adiabatic change of the laser phase $\varphi $ by $\pi
$ will result in a $\tilde{\sigma}_{z}$ quantum gate operation. These
operations are expected to be used for manipulating the quantum information
stored in the topological quantum memory.

We now consider the detection of the non-Abelian braiding statistics of
Majorana fermions. It is known that, if one exchanges the Majorana fermion $%
\gamma _{i}$ with its nearest neighbor one $\gamma _{i+1}$, one can realize
the braiding operation $\tau _{i}=\exp (\frac{\pi }{4}\gamma _{i+1}\gamma
_{i})$ {\color{blue}\cite{Ivanov}}. The non-Abelian braiding statistics
arises from the fact that $\tau _{i}\tau _{i+1}\neq \tau _{i+1}\tau _{i}$.
So in order to demonstrate this non-Abelian braiding statistics, we at least
need four Majorana fermions. As shown in Fig. 2(c), by addressing some
particular optical lattice sites, the two optical lattice segments $L$ and $R
$ are driven into topological phases and two pairs of Majorana fermions $%
\left( \gamma _{1},\gamma _{2}\right) $ and $\left( \gamma _{3},\gamma
_{4}\right) $ are created at their ends. Based on these Majorana end modes,
we define two Dirac fermions $c_{L}=$ $(\gamma _{1}+i\gamma _{2})/2$ and $%
c_{R}=$ $(\gamma _{3}+i\gamma _{4})/2$. In Fig. 2(d), as the way in the
superconducting wire model {\color{blue}\cite{Alicea}}, we introduce a cross
lattice and use the basic operations in Fig. 2(a-b) to braid the Majorana
fermions at the ends of the same or different topological segments, which
correspond to the braiding operations $\tau _{1(3)}$ and $\tau _{2}$.
Following the experimental method in {\color{blue}\cite{Weitenberg}}, as
shown in Fig. 1(c), the braiding operations involving two orthogonal optical
lattice segments can be realized by using a 2D Mott insulator with unitary
filling. Initially the atoms in the 2D Mott state are in a spin state $|\chi
\rangle \neq |\sigma \rangle $. Using single spin addressing, the two
orthogonal lattice segments are created in the $|\sigma \rangle $. If $%
V_{\chi ,\sigma }\gg V_{\sigma ,\sigma ^{\prime }}$, the dynamic is frozen
between $|\chi \rangle $ and $|\sigma \rangle $ neighboring sites. Here we
assume the lattice has simulated the transverse field $\sigma ^{x}$-Ising
model. In the fermionic occupation basis $\left\{ \left\vert 00\right\rangle
_{LR}\text{, }c_{L}^{+}\left\vert 00\right\rangle _{LR}\text{, }%
c_{R}^{+}\left\vert 00\right\rangle _{LR}\text{, }c_{L}^{+}c_{R}^{+}\left%
\vert 00\right\rangle _{LR}\right\} $, the two braiding operations $\tau _{1}
$ and $\tau _{2}$ are represented by
\begin{equation}
\tau _{1}=\frac{1}{\sqrt{2}}\text{diag}(1-i,1+i,1-i,1+i),
\end{equation}

\begin{equation}
\tau _{2} =\frac{1}{\sqrt{2}}\left(
\begin{array}{cccc}
1 &  &  & -i \\
& 1 & -i &  \\
& -i & 1 &  \\
-i &  &  & 1%
\end{array}%
\right) .
\end{equation}


Based on these operations, we construct two composite braiding operations to
demonstrate $\tau _{1}\tau _{2}\neq \tau _{2}\tau _{1}$. The two composite
braiding operations are chosen as $S=\tau _{1}\tau _{2}$, $T=\tau _{2}\tau
_{1}$ with the properties


\begin{equation}
ST=-i\tilde{\sigma}_{x}^{L}\otimes \tilde{\sigma}_{x}^{R}\text{, \ }TS=-i%
\tilde{\sigma}_{z}^{L}\otimes I^{R}.
\end{equation}%
Suppose the two Majorana fermions are initially prepared in the state $%
\left( \left\vert 0\right\rangle +\left\vert 1\right\rangle \right)
_{L}\otimes \left( \left\vert 0\right\rangle +\left\vert 1\right\rangle
\right) _{R}/2$, after the braiding operations $ST$ and $TS$, the output
states $-i\left( \left\vert 0\right\rangle +\left\vert 1\right\rangle
\right) _{L}\otimes \left( \left\vert 0\right\rangle +\left\vert
1\right\rangle \right) _{R}/2$ and $-i\left( \left\vert 0\right\rangle
-\left\vert 1\right\rangle \right) _{L}\otimes \left( \left\vert
0\right\rangle +\left\vert 1\right\rangle \right) _{R}/2$ are orthogonal
with each other. Thus we can demonstrate the non-Abelian nature of the
Majorana fermions unambiguously by detecting the difference of the two
orthogonal output states.

In fact, this detection is readily available in experiment by transferring
them into spin basis. In this basis, the two degenerate ground states of the
topological lattice segment $i$ are written as

\begin{equation}
\left\vert 0\right\rangle _{i}=\frac{\left\vert +\right\rangle ^{\otimes
m_{i}}+\left\vert -\right\rangle ^{\otimes m_{i}}}{\sqrt{2}}\text{, }%
\left\vert 1\right\rangle _{i}=\frac{\left\vert +\right\rangle ^{\otimes
m_{i}}-\left\vert -\right\rangle ^{\otimes m_{i}}}{\sqrt{2}},
\end{equation}%
where $\left\vert \pm \right\rangle =(\left\vert \uparrow \right\rangle \pm
\left\vert \downarrow \right\rangle )/\sqrt{2}$. The Dirac fermion and the
fermion parity operators become $c_{i}=(\sigma _{1}^{x}-\sigma
_{m_{i}}^{x}\prod_{j=1}^{m_{i}}\sigma _{j}^{z})/2$ and $P_{i}=%
\prod_{j=1}^{m_{i}}\sigma _{j}^{z}$, where $i=L,R$, $m_{i}$ is the total
lattice sites in the topological segment $i$. Using these expressions, one
can demonstrate $c_{i}^{+}\left\vert 0\right\rangle _{i}=\left\vert
1\right\rangle _{i}$, $c_{i}\left\vert 1\right\rangle _{i}=\left\vert
0\right\rangle _{i}$ and $P_{i}\left\vert 0\right\rangle _{i}(\left\vert
1\right\rangle _{i})=\left\vert 0\right\rangle _{i}(-\left\vert
1\right\rangle _{i})$ as in the JW fermions basis. By substituting Eq. (10),
the initial state of Majorana fermions is transformed into $\left\vert
+\right\rangle ^{\otimes m_{L}}\left\vert +\right\rangle ^{\otimes m_{R}}$,
which can be easily prepared by optical pumping, and the two output states
become $-i\left\vert +\right\rangle ^{\otimes m_{L}}\left\vert
+\right\rangle ^{\otimes m_{R}}$ and $i\left\vert -\right\rangle ^{\otimes
m_{L}}\left\vert +\right\rangle ^{\otimes m_{R}}$, which can be
adiabatically rotated into $-i\left\vert \uparrow \right\rangle ^{\otimes
m_{L}}\left\vert \uparrow \right\rangle ^{\otimes m_{R}}$ and $i\left\vert
\downarrow \right\rangle ^{\otimes m_{L}}\left\vert \uparrow \right\rangle
^{\otimes m_{R}}$ by another laser driving. That is, we only need to
distinguish the two orthogonal collective spin states $\left\vert \downarrow
\right\rangle ^{\otimes m_{L}}$ and $\left\vert \uparrow \right\rangle
^{\otimes m_{L}}$. For this purpose, one can apply one laser beam on the
lattice segment $L$ to couple the spin up state to an auxiliary excited
state and detect the fluorescence of its emissions. The cases with and
without fluorescence are corresponding to the state $\left\vert \uparrow
\right\rangle $ and $\left\vert \downarrow \right\rangle $. The distinct
advantage here is that the state $\left\vert \uparrow \right\rangle
^{\otimes m_{L}}$ is a collective spin state which can provide a strong
fluorescence to distinguish itself from the state $\left\vert \downarrow
\right\rangle ^{\otimes m_{L}}$. Such property makes our proposal much more
appealing for the detection the non-Abelian statistics of Majorana fermions.

In summary, we have proposed an experimental scheme to create and manipulate
Majorana fermions. In the scheme, the transverse field Ising model has been
simulated using cold atoms trapped in optical lattice. By tuning the
intensity of the addressing lasers, Majorana fermions can be generated at
the ends of the topological lattice segment. Such tuning can also allow them
to be moved, fused and braided. Finally, we have shown how to detect the
non-Abelian statistics of Majorana fermions by distinguishing two orthogonal
output collective spin states immediately after performing two opposite
orderings of braiding operations.

\bigskip

\noindent \textit{Acknowledgments -} F. Mei thanks Dongling Deng and J.
Alicea for many helpful discussions. This work was supported by the NSFC
under Grant No. 60978009, No. 11125417 and No. 11074079, the Major Research
Plan of the NSFC (No. 91121023), the SKPBR of China ( No.2009CB929604,
No.2011CB922104 and 2011CBA00200), the NUS Academic Research (Grant No. WBS:
R-710-000-008-271), the Ministry of Education of Singapore and the
Postgraduate Scholarship of China Scholarship Council.

\bigskip

\bigskip

\bigskip

\bigskip

\bigskip

\

\bigskip

\bigskip

\bigskip

\bigskip

\bigskip

\bigskip

\bigskip

\bigskip

\bigskip

\bigskip

\

\bigskip

\bigskip

\bigskip

\bigskip

\bigskip

\bigskip


\begin{thebibliography}{99}
\bibitem{Wilczek} F. Wilczek, Nature Phys. \textbf{5}, 614 (2009).

\bibitem{TQC} C. Nayak \textit{et al.,} Rev. Mod. Phys. \textbf{80}, 1083
(2008).

\bibitem{Read} N. Read and D. Green, Phys. Rev. B \textbf{61}, 10267 (2000).

\bibitem{Kitaev01} A. Yu. Kitaev, Phys. Usp. \textbf{44}, 131 (2001).

\bibitem{2DMF} J.D. Sau \textit{et al.} Phys. Rev. Lett. \textbf{104},
040502 (2010); J. Alicea, Phys. Rev. B \textbf{81}, 125318 (2010).

\bibitem{1DMF} R.M. Lutchyn, J.D. Sau, and S. Das Sarma, Phys. Rev. Lett.
\textbf{105}, 077001 (2010); Y. Oreg, G. Refael, and F. von Oppen, Phys.
Rev. Lett. \textbf{105}, 177002 (2010).

\bibitem{Alicea} J. Alicea \textit{et al.,} Nature Phys. \textbf{7}, 412
(2011).

\bibitem{Spielman1} Y.-J. Lin \textit{et al.}, Nature (London) \textbf{462},
628 (2009); M. Aidelsburger \textit{et al.}, Phys. Rev. Lett. \textbf{107},
255301 (2011).

\bibitem{Spielman2} Y.-J. Lin \textit{et al.}, Nature (London) \textbf{471},
83 (2011).

\bibitem{Zhu} S. L. Zhu \textit{et al.}, Phys. Rev. Lett. \textbf{106},
100404 (2011).

\bibitem{Tewari} S. Tewari \textit{et al.}, Phys. Rev. Lett. \textbf{98},
010506 (2007).

\bibitem{Santo} M. Sato, Y. Takahashi and S. Fujimoto, Phys. Rev. Lett.
\textbf{103}, 020401 (2009).

\bibitem{Liu} X. J. Liu \textit{et al.}, arXiv:1111.1798v1.

\bibitem{Jiang} L. Jiang \textit{et al.}, Phys. Rev. Lett. \textbf{106},
220402, (2011).

\bibitem{Zoller} C. V. Kraus \textit{et al.}, arXiv:1201.3253; C.-E. Bardyn
\textit{et al.}, arXiv:1201.2112.

\bibitem{Kitaev09} A. Kitaev and C. Laumann, arXiv:0904.2771.

\bibitem{You-Solano} J. Q. You \textit{et al.}, arXiv:1108.3712; A.
Mezzacapo \textit{et al.}, arXiv:1108.3712.

\bibitem{Duan} L.-M. Duan, E. Demler and M.D. Lukin, Phys. Rev. Lett.
\textbf{91}, 090402, (2003).

\bibitem{Mott02} M. Greiner \textit{et al.}, Nature (London) \textbf{415},
39 (2002).

\bibitem{Bakr} W.S. Bakr \textit{et al.}, Nature (London) \textbf{462, } 74
(2009).

\bibitem{Sherson} J.F. Sherson \textit{et al.}, Nature (London) \textbf{467,
} 68 (2010).

\bibitem{Ivanov} D.A. Ivanov, Phys. Rev. Lett. \textbf{86}, 268, (2001).

\bibitem{Weitenberg} C. Weitenberg \textit{et al.}, Nature (London) \textbf{%
471, } 319 (2011).
\end{thebibliography}
\end{document}